\documentclass[aps,twocolumn,twoside, nofootinbib, superscriptaddress]{revtex4}

\usepackage{amsthm}
\usepackage{amsmath,latexsym,amssymb,verbatim,enumerate}

\usepackage{graphicx}
\usepackage{color}

\newcommand{\ket}[2][]{{|#2\rangle_{#1}}}
\newcommand{\bra}[2][]{{}_{#1}\langle #2|}

\def\be{\begin{equation}}
\def\ee{\end{equation}}
\def\ben{\begin{eqnarray}}
\def\een{\end{eqnarray}}
\def\bei{\begin{itemize}}
\def\eei{\end{itemize}}

\def\<{\langle}
\def\>{\rangle}

\def\<{\langle}
\def\>{\rangle}
\def\tvec{\textbf}
\def\tsym{\boldsymbol}

\begin{document}

\title{Redundant information encoding in QED during decoherence}
\author{J.~Tuziemski}
\affiliation{Faculty of Applied Physics and Mathematics, Gda\'nsk University of Technology, 80-233 Gda\'nsk, Poland}
\affiliation{National Quantum Information Centre in Gda\'nsk, 81-824 Sopot, Poland}
\author{P.~Witas}
\affiliation{Institute of Physics, Faculty of Physics, Astronomy and Informatics
	Nicolaus Copernicus University, 87-100 Toru\'n, Poland}
\author{J.~K.~Korbicz}
 \email{jkorbicz@mif.pg.gda.pl}
  \affiliation{Faculty of Applied Physics and Mathematics, Gda\'nsk University of Technology, 80-233 Gda\'nsk, Poland}
 \affiliation{National Quantum Information Centre in Gda\'nsk, 81-824 Sopot, Poland}

\begin{abstract}
Broadly understood decoherence processes in quantum electrodynamics, induced by neglecting either the radiation [L. Landau, Z. Phys. {\bf 45}, 430 (1927)] 
or the charged matter [N. Bohr and L. Rosenfeld, K. danske vidensk. Selsk, Math.-Fys. Medd. {\bf XII}, 8 (1933)],
have been studied from the dawn of the theory. 
However what happens in between, when a part of the radiation may be observed, as is the case in many real-life situations, 
has not been analyzed yet.  We present such an analysis for  a non-relativistic, point-like charge and thermal radiation.
In the dipole approximation, we solve the dynamics and  
show that there is a regime where, despite of the noise, the observed 
field carries away almost perfect and hugely redundant information about the charge momentum. 
We analyze a partial charge-field state and show that it approaches a so called spectrum broadcast structure.

\end{abstract}

\maketitle
\section{Introduction}
Quantum information theory approach to open quantum systems has been a subject of an active research recently,
with an advent of such new and exciting research areas as thermodynamics of meso- and nanoscale 
systems \cite{michal,popescu,2ndlaw} and quantum Darwinism \cite{ZurekNature, dec, phystoday}, to name just two.
Here we consider quantum electrodynamics (QED) from an open system's perspective ( see \cite{BPbook,Joos,RelQED,BCP2006} and the references therein),
treating the electromagnetic field as the environment for the charge. We use quantum information concepts to
study information gained by portions of the (initially thermal) field about the charge during the evolution. Consequently, we have to go beyond the usual 
approach to open systems, where the environment is assumed to pass unobserved and hence is traced out, and only the reduced state of the system is explicitly
studied (see e.g. \cite{dec,BPbook}).
This leads, under appropriate conditions, to the well known phenomenon of decoherence, i.e. the loss of coherences
in some preferred basis of the system, called the pointer basis. This phenomenon has been experimentally 
observed in a variety of systems \cite{decohexp}. In QED, decoherence due to various effects has been extensively studied (see e.g. \cite{Joos} for a review),
including decoherence due to the dressing (e.g. in \cite{BCP2006}), non-zero temperature \cite{BCP2006}, Bremsstrahlung \cite{RelQED,BPbook},
and the charge monitoring the field \cite{Joos}. 

Here instead, we assume that a part of the field is monitored and thus cannot be traced out. This line of thinking has been introduced
in the quantum Darwinism program \cite{ZurekNature, dec, phystoday} and further developed in the, so called, spectrum broadcast structure approach to objectivity
\cite{pra,sfera,qbm,photonics} (see also \cite{Pawel}). In the spirit of the latter, we study a partially traced state, containing 
a part of the radiation modes.
We show that under appropriate conditions and a certain coarse-graining, almost perfect information about the charge momentum is 
encoded during the decoherence into the thermal field with a huge redundancy. 
It can be in principle extracted via projective measurements on the field modes with negligible disturbance to the 
partial charge-field state. This result is achieved via showing that the partially traced state approaches the, so called,
spectrum broadcast structure (SBS) \cite{pra, sfera, qbm, photonics}---a state structure describing broadcasting of the same classical
information into multiple quantum systems.
Some preliminary results along these lines were obtained in \cite{BCP2006}, where a build up of correlations between momentum components of the charge 
and the dressing cloud was shown during the vacuum induced decoherence.  But neither the structure of the partially traced state
has been considered nor the redundancy of information shown. Also, we show the redundant information transfer for the
thermal, rather than for the vacuum field, which is more realistic and surprising due to the inherent noise.

We consider the non-relativistic regime of QED and neglect any possible inner degrees of freedom of the charge, 
treating it as a free, point-like particle of mass $m_0$ and charge $q$, interacting with initially thermal field. 
The charge-field system is then described by the 
minimally coupled Hamiltonian with a necessary cut-off frequency $\bar\Omega$ to avoid the ultraviolet divergences:
\be
\hat{H} = \frac{1}{2m_0} \left[\hat{\tvec p} - q \hat{\tvec A}(\hat {\tvec r}) \right]^2 + \sum_{\tvec{k},j} \hbar \omega_\tvec{k} \hat{a}^\dagger_{\tvec{k},j} \hat{a}_{\tvec{k},j},
\ee 
where the potential $\tvec A({\tvec r})$ is chosen in the Coulomb gauge:
\ben\label{A}
\hat{\tvec A}(\hat{\tvec r}) = \sum_{\tvec{k},j} \tsym{\epsilon}_{\tvec{k},j}\sqrt \frac{ \hbar}{2\varepsilon_0 \omega_\tvec{k} V}\left(\hat{a}^\dagger_{\tvec{k},j}e^{-i \tvec{k} \cdot \hat{\tvec r}} +\hat{a}_{\tvec{k},j}e^{\tvec{k} \cdot \hat{\tvec r}} \right).
\een 
Here $\tsym{\epsilon}_{\tvec{k},j}$ is the polarization vector of the mode $\tvec k$, $\omega_\tvec{k}$ is its frequency, the field is quantized in a box of volume $V$
with the sum restricted to $\omega_\tvec k\lesssim\bar\Omega$, 
and $\hat{a}_{\tvec{k},j},\hat{a}^\dagger_{\tvec{k},j}$ are the creation and anihilation operators obeying $[\hat{a}_{\tvec{k},j}, \hat{a}^\dagger_{\tvec{k}',j'}] = \delta_{\tvec{k},\tvec{k}'} \delta_{j,j'}$. We consider the charge initially described by a wave packet localized at $\delta r_0$ small 
with respect to the shortest relevant wavelength of the field and the cut-off is assumed to reflect this. 
The spreading will limit the use in such cases of the usual
dipole approximation (see e.g. \cite{Sakurai})  to times not much larger than $\bar\Omega^{-1}$.
To somewhat improve the situation, we will use the, so called, moving dipole approximation,
introduced in \cite{BCP2006} and giving longer times.
One follows with the dipole approximation the average packet position, assumed to travel along the free   
trajectory $\tvec r(t)=\tvec r_0+\tvec v_0 t$, with $\tvec v_0$ the initial average particle velocity. 
The approximation breaks down when the packet width
becomes comparable with $c/\bar\Omega$, which happens for:
\be\label{tdip}
t\lesssim\tau_{dip} \equiv \frac{m_0 c}{\bar\Omega \delta p_0},
\ee 
obtained assuming minimal initial packet and free (i.e. non-interacting) spreading. The consistency of this approximation has been proven
in \cite{BCP2006}. Due to (\ref{tdip}), the information accumulation effects 
  require, as we will see, very strong coupling, but, nevertheless, are in principle possible. Since we are interested in moderate field intensities we  neglect the  $\hat{\tvec A}(\tvec{r}(t))^2$ term, which leads to \cite{Sakurai}:

\ben\label{H}
\hat{H}\approx\frac{\hat{\tvec p}^2}{2m_0} + \sum_{\tvec{k},j} \hbar \omega_\tvec{k} \hat{a}^\dagger_{\tvec{k},j} \hat{a}_{\tvec{k},j} - \frac{q}{m_0} \hat{\tvec p} \cdot \hat{\tvec A}(\tvec{r}(t)).
\een

\begin{figure}
\centering
\includegraphics[scale=0.3]{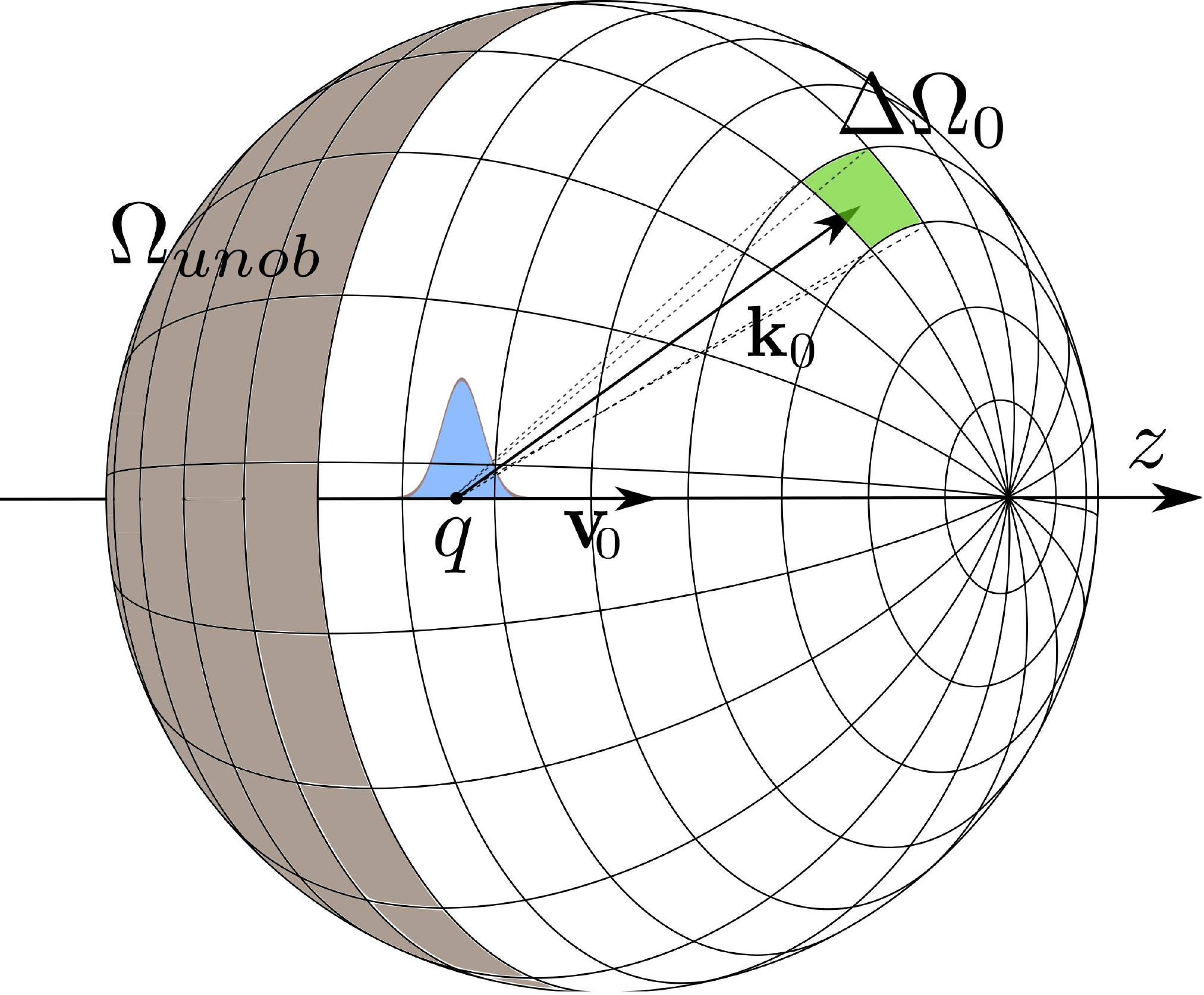}
\caption{(Color online). The considered physical model: A non-relativistic charged particle 
interacts with electromagnetic field, treated as the environment. The particle is described by a wave packet, 
narrow compared to the shortest relevant radiation wavelength, and moves with initial velocity $\tvec v_0$, chosen along the $z$-axis.
The sphere represents the "celestial sphere" of the mode directions $\tvec k/k$. A part of this sphere, given by a solid angle $\Omega_{unob}$,
is not monitored and the corresponding modes are traced out. The rest is divided into small portions (only one portion shown)
$\Delta\Omega_0$, centered each around some average direction $\tvec k_0/k_0$, which represent e.g. detection regions of approximately point-like
detectors. 
}\label{fig1}
\end{figure}

\section{Calculation of the partially traced state}
Our main object of the study is a partially traced state, with a part of the field included in the description:
\be\label{rhoobs}
\varrho_{S:F_{obs}}(t)\equiv tr_{F_{unob}}[U_{S:F}(t)\varrho_0U_{S:F}(t)^\dagger], 
\ee
where $S$ is the charge, and $F_{obs}$, $F_{unob}$ denote the observed and unobserved modes 
respectively and $U_{S:F}(t)$ is the evolution operator corresponding to (\ref{H}).
The latter can be found exactly (cf. Eqs. (8,A12) in \cite{BCP2006}; a similar derivation can be found also in \cite{qbm}) and is given in the interaction picture by:
\ben\label{U}
\hat{U}^{I}_{S:F}(t) = \int d^3 p \ket{\tvec{p}}\bra{\tvec{p}} \otimes \hat{U}^{I}_{F}(t;\tvec{p}), 
\een
which is a controlled-unitary type of evolution \cite{sfera}  (see also \cite{ZQZ2009,ZQZ2010,ZRZ2012}) with:
\ben
\hat{U}^{I}_{F}(t;\tvec{p})\equiv &&e^{i \sum_{\tvec{k},j} C_{\tvec{k}} \tvec{p} \cdot \tsym{\epsilon}_{\tvec{k},j}\xi_\tvec{k}(t)}\times\nonumber\\
&&\times\hat{D}\Big( \sum_{\tvec{k},j} C_{\tvec{k}}\tvec{p} \cdot \tsym{\epsilon}_{\tvec{k},j}  \alpha_{\tvec{k}}(t)\Big).\label{Up} 
\een
where $\hat{D}(\sum_{\tvec{k},j} \beta_{\tvec{k},j}) \equiv\exp[ \sum_{\tvec{k},j} (\beta_{\tvec{k},j} \hat{a}_{\tvec{k},j}^{\dagger}- \beta_{\tvec{k},j}^* \hat{a}_{\tvec{k},j})]$ \
is the multi-mode displacement operator, $C_{\tvec{k}} \equiv - (q/m_0)\sqrt{\hbar/(2 \varepsilon_0 \omega_\tvec{k} V)}$ is a coupling coefficient,
\be\label{ak}
\alpha_{\tvec{k}}(t) \equiv e^{-i\tvec{k} \cdot \tvec{r}_0}\frac{1-e^{i(\omega_\tvec{k}- \tvec{k} \cdot \tvec{v}_0)t}}{\hbar\left(\omega_{\tvec k} - \tvec{k} \cdot \tvec{v}_0\right)},
\ee
and $\xi_\tvec{k}(t)\equiv [t - \sin(\tvec{k} \cdot \tvec{v}_0t)/(\tvec{k} \cdot \tvec{v}_0)]/(\omega_\tvec{k}- \tvec{k} \cdot \tvec{v}_0)$ is a dynamical phase, which will turn out to be irrelevant for our considerations. 
Note also that, since $\hat{\tvec p}$ commutes with (\ref{H}), the momentum of the charge is conserved during 
the evolution in the dipole approximation, so that in particular the momentum spread is constant in time.

Following the standard approach, the charge-field system is assumed to be initially in a product state \cite{Sakurai, BCP2006}:
\be\label{init}
\varrho_0=\ket{\psi_{0S}}\bra{\psi_{0S}}\otimes\varrho_{0F},
\ee 
where $\ket{\psi_{0S}}$ is a charge initial wave packet and the field is in a thermal state, $\varrho_{0F}=\exp(-\beta \hat H_F)/Z(\beta)$, 
$\hat H_F\equiv \sum_{\tvec{k},j} \hbar \omega_\tvec{k} \hat{a}^\dagger_{\tvec{k},j} \hat{a}_{\tvec{k},j}$, 
$\beta\equiv \hbar /k_BT$. 
This, to some extent artificially, decoupled state 
leads at the very short time-scale $t\sim\bar\Omega^{-1}$ to the well known effects of dressing  and charge energy renormalization \cite{Sakurai, BCP2006}.
To separate those transient effects from the thermal influence, in what follows we assume the low thermal energy regime \cite{BCP2006}: 

\be
k_BT\ll \hbar\bar\Omega.
\ee

The spreads of the initial wave packet $\ket{\psi_{0S}}$ are assumed to satisfy $\delta r_0\ll c/\bar\Omega$ and 
obviously $\delta p_0\ll m_0c$,
which warrants the moving dipole approximation for times (\ref{tdip}) $\tau_{dip}\gg\bar\Omega^{-1}$.
However, as will be clear later, $\delta p_0$ cannot be chosen too small either. 

Under the above conditions, one can find the partially traced state (\ref{rhoobs}) using (\ref{U},\ref{Up}). 
Although (\ref{U}) is formally written with the integral and the sharp momentum eigenstates, one should keep in mind that 
by the spectral theorem it is in fact a limit over finite divisions $\left\{ \Delta\right\}$ of the momentum space $\mathbf R^3$, of sums with $\left|\tvec{p}\right\rangle \left\langle  \tvec{p} \right|$ approximated 
by the spectral projectors $\hat{\Pi}_{\Delta}$. We thus obtain  that in the interaction picture:
\ben
\label{mama}
&&\varrho_{S:F_{obs}}^I(t) \!=\!\!
\sum_{\Delta} \hat{\Pi}_{\Delta} \varrho^{I}_{0S} \hat{\Pi}_{\Delta} \otimes \varrho^{I}_{F_{obs}}(t;\tvec{p}_{\Delta}) +
\!\!\!\! \sum_{\Delta \neq \Delta' } D_{\tvec{p}_{\Delta}\!,\tvec{p}_{\Delta'}} \nonumber\\
&& \times\hat{\Pi}_{\Delta} \varrho^{I}_{0S} \hat{\Pi}_{\Delta'} 
\otimes \hat{U}^{I}_{F_{obs}}(t;\tvec{p}_{\Delta}) \varrho^{I}_{0F_{obs}} \hat{U}^{I}_{F_{obs}}(t;\tvec{p}_{\Delta'})^{\dagger},  
\een         
where $\tvec{p}_{\Delta}$ is some point from $\Delta$, $\varrho_{0F_{obs}}\equiv tr_{unob}\varrho_{0F}$, 
$\hat{U}_{F_{obs}}(t;\tvec{p})\equiv tr_{unob}\hat{U}_{F}(t;\tvec{p})$ (cf. (\ref{Up})), and:
\ben
&&\varrho^{I}_{F_{obs}}(t;\tvec{p}) \equiv \hat{U}^{I}_{F_{obs}}(t;\tvec{p}) \varrho_{0F_{obs}} \hat{U}^{I}_{F_{obs}}(t;\tvec{p})^{\dagger}, \label{Fobs}\\ 
&&D_{\tvec{p}, \tvec{p}'}(t)\equiv tr [\hat{U}_{F_{unob}}(t;\tvec{p}) \varrho_{0F_{unob}} \hat{U}_{F_{unob}}(t;\tvec{p}')^{\dagger}] \\
&&\equiv\exp\left[-\Gamma_{\tvec{p}, \tvec{p}'}(t)+i\Phi_{\tvec{p}, \tvec{p}'}(t)\right] \label{eq:decf}
\een        
the latter being the decoherence factor due to the unobserved field modes (the same in the interaction and the Schr\"odinger pictures). 
The real part $\Gamma_{\tvec{p}, \tvec{p}'}(t)$ leads to
the damping of coherences in the momentum basis and singles it out as the pointer basis. 
The resulting suppression of the charge-field entanglement is a necessary condition
for the appearance of objectivity \cite{sfera, pra}.

\section{Decoherence processes}
The decoherence process in this model has been extensively studied in \cite{BCP2006} 
with the whole of the radiation traced out.  The results can be easily generalized to our situation
where only a portion $F_{unob}$ of the modes is neglected. We assume it is macroscopic, i.e. contains large enough number of modes
to pass to the continuum limit $\sum_{\tvec{k}}\to V\int_{F_{unob}} d^3 k / (2 \pi)^3$, where $F_{unob}$
is described by an angle $\Omega_{unob}$ of the unobserved directions (see Fig.~\ref{fig1}), containing all the relevant frequencies and polarizations. 
Using $k_BT\ll\hbar\bar\Omega$ and $v_0/c\ll1$ we obtain:
\ben
&&\frac{\pi}{\alpha} \Gamma_{\tvec{p}, \tvec{p}'}(t)=\nonumber\\
&&=\left[F_0(\Delta \tvec{p})+\frac{v_0}{c}F_1(\Delta \tvec{p})\right]\log \left[\sqrt{1+\bar\Omega^2 t^2} \frac{\sinh\left(t/\tau_F\right)}{t/\tau_F} \right]\nonumber \\ 
&&-\frac{v_0}{2c}F_1(\Delta \tvec{p})\left[\frac{t}{\tau_F}\coth\frac{t}{\tau_F} -\frac{1}{1+ \bar\Omega^2 t^2}\right]+O\left(\frac{v_0^2}{c^2}\right),\label{logD}
\een
where $\alpha\equiv q^2/(4 \pi \varepsilon_0 \hbar c)$ is proportional to the fine structure constant, $\tau_F \equiv \hbar/(\pi k_B T)$ 
is the characteristic thermal time, $\Delta\tvec p\equiv \tvec p-\tvec p'$, and:  
\ben
&&F_0(\Delta \tvec{p})\equiv \frac{1}{(m_0 c)^{2}}\int_{\Omega_{unob}}\frac{d\Omega_{\tvec k}}{4\pi}\Delta \tvec{p}^2_{\perp\tvec k},\label{F0}\\
&&F_1(\Delta \tvec{p})\equiv\frac{2}{(m_0 c)^{2}}\int_{\Omega_{unob}}\frac{d\Omega_{\tvec k}}{4\pi}\cos\theta_\tvec k\Delta \tvec{p}^2_{\perp\tvec k}.
\label{F1}\een 
Here $F_0(\Delta \tvec{p}), F_1(\Delta \tvec{p})$ are the average and the "first moment" 
of the squared norm of the transversal part of 
$\Delta \tvec{p}$, $\Delta \tvec{p}^2_{\perp\tvec k}\equiv\sum_{ij}\Delta \tvec{p}_i\Delta \tvec{p}_j(\delta_{ij}-k_ik_j/k^2)$,
over the unobserved 
directions $\Omega_{unob}$ and rescaled to $(m_0 c)^2$. 
A comment is in order. The quantities (\ref{F0},\ref{F1}) are formally second order in $1/c$. This is however not a mismatch in the
relativistic expansion as it may first appear due to non-relativistic Hamiltonian used. This is rather a result of the continuum limit and 
the wave nature of light as we illustrate in more detail in Appendix \ref{rel_cor}.
 
Since generically $F_1(\Delta \tvec{p})\ne 0$, there is in general a non-vanishing first order contribution
to the decoherence factor from the Doppler shift (cf. (\ref{ak})). If however all 
of the field is neglected,  $F_1(\Delta \tvec{p})=0$ \cite{BCP2006} and in the first order (\ref{logD})
is the same as for a static wave packet ($\tvec v_0=0$).
One easily sees from (\ref{logD}) that the decoherence factor depends on the time via $\bar\Omega t$ and $t/\tau_F$.
This defines three time-dependence regimes, with the following approximate behavior in each of them \cite{BCP2006}:
\ben
&&\frac{\pi}{\alpha} \Gamma_{\tvec{p}, \tvec{p}'}(t)\approx \label{Gappr}\\
&&\left\{\begin{array}{l r} F_0\frac{\bar\Omega^2t^2}{2}, & t\ll \bar\Omega^{-1}\\
\left(F_0+\frac{v_0}{c}F_1\right)\log \bar\Omega t-\frac{v_0}{2c}F_1, & \bar\Omega^{-1}\ll t\ll\tau_F\\
\left(F_0+\frac{v_0}{2c}F_1\right)\frac{t}{\tau_F}+\left(F_0+\frac{v_0}{c}F_1\right)\log \bar\Omega\tau_F, & 
t\gg\tau_F.
\end{array} \right.\nonumber
\een
The initial "vacuum decoherence" for $t\ll\tau_F$, accompanying the dressing and the mass renormalization \cite{BCP2006}, 
is a consequence of the artificially decoupled initial state (\ref{init}). Past this transient period, for 
$t\sim\tau_F$ the thermally driven decoherence begins, giving the exponential decay of coherences with time.
Since $\tau_{dip}/\tau_F=(m_0c/\delta p_0)(k_BT/\hbar\Omega)$, one can achieve  $\tau_F<\tau_{dip}$ in the studied regime
so that it can be in principle observed within the dipole approximation. However, while the fundamental time limit 
(\ref{tdip}), imposed by the wave packet spread, grows linearly with $m_0c/\delta p_0$, 
the decoherence factor decays only as $|D_{\tvec p,\tvec p'}|\sim\exp[-\alpha(m_0c/\delta p_0)^{-2}]$, since
from (\ref{F0},\ref{F1}), $F_0(\Delta \tvec{p}),F_1(\Delta \tvec{p})\sim(m_0c/\delta p_0)^{-2}$.
Thus, what is required is a not-so-small momentum spread 
and a very strong coupling, $\alpha\gg 1$, corresponding to macroscopic charges.  
A sample plot of such a situation is shown in Fig.~\ref{dec}.

\section{Information content of the radiation field}
We now move to the most interesting part---the information content of the observed radiation modes (\ref{Fobs}),
which has not been studied explicitly in this model.
Let us first look at an individual mode $\tsym\epsilon_{\tvec k , j}$. From (\ref{Up},\ref{mama}) its state is a mixture
of displaced initial thermal states:
\be\label{rhomic}
\varrho^I_{\tvec k,j}(t;\tvec p)
\equiv \hat{D}_{\tvec{p} \cdot \tsym{\epsilon}_{\tvec{k},j}}(t)\varrho_{0\tvec k,j}\hat{D}_{\tvec{p} \cdot \tsym{\epsilon}_{\tvec{k},j}}(t)^\dagger,
\ee
where $\hat{D}_{\tvec{p} \cdot \tsym{\epsilon}_{\tvec{k},j}}(t)$ stands for each of the  displacements in (\ref{Up}).
These displacements depend on the component of the charge momentum along the mode polarization and 
we can ask how distinguishable are two such states for different $\tvec p\cdot \tsym{\epsilon}_{\tvec{k},j}$.
As the appropriate measure, we choose  the mixed state fidelity ( also known as generalized overlap) $B(\varrho, \sigma) \equiv tr\sqrt{\sqrt{\varrho}\sigma \sqrt{\varrho}}$ \cite{Fuchs,sfera}, 
satisfying $B(\varrho, \sigma)=0$ if and only if $\varrho$ and $\sigma$ have orthogonal supports and hence are perfectly distinguishable.
$B[\varrho_{\tvec k,j}(t;\tvec p),\varrho_{\tvec k,j}(t;\tvec p')]\equiv B_{\tvec{p}, \tvec{p}'}^{(\tvec{k},j)}(t)$ 
can be calculated, using e.g. the techniques of \cite{qbm} and reads \cite{nota2}:
\ben
&&\log B_{\tvec{p}, \tvec{p}'}^{(\tvec{k},j)}(t) = -\frac{\alpha \pi\hbar^2 c(\tsym\epsilon_{\tvec k , j}\cdot\Delta\tvec{p})^2}{m_0^2\omega_\tvec k V} \left|\alpha_{\tvec{k}}(t)\right|^2 
\tanh \left(\frac{\beta \omega_k}{2} \right). \nonumber \\&&\label{Bmic}
\een
If $\tsym\epsilon_{\tvec k , j}\cdot\Delta\tvec{p}\ne 0$, (\ref{ak}) implies that it oscillates with a Doppler-shifted frequency 
$\omega_\tvec k[1-\tvec k\cdot\tvec v _0/(kc)]$.
However, in the infrared limit $V\to \infty$, $B_{\tvec{p}, \tvec{p}'}^{(\tvec{k},j)}(t) \to 1$, indicating that the states (\ref{rhomic}) 
become identical for all $\tvec p$. Thus, on the microscopic level each field mode carries vanishingly small information about the charge (cf. \cite{sfera}). 

\begin{figure}
\centering
\includegraphics[scale=0.35]{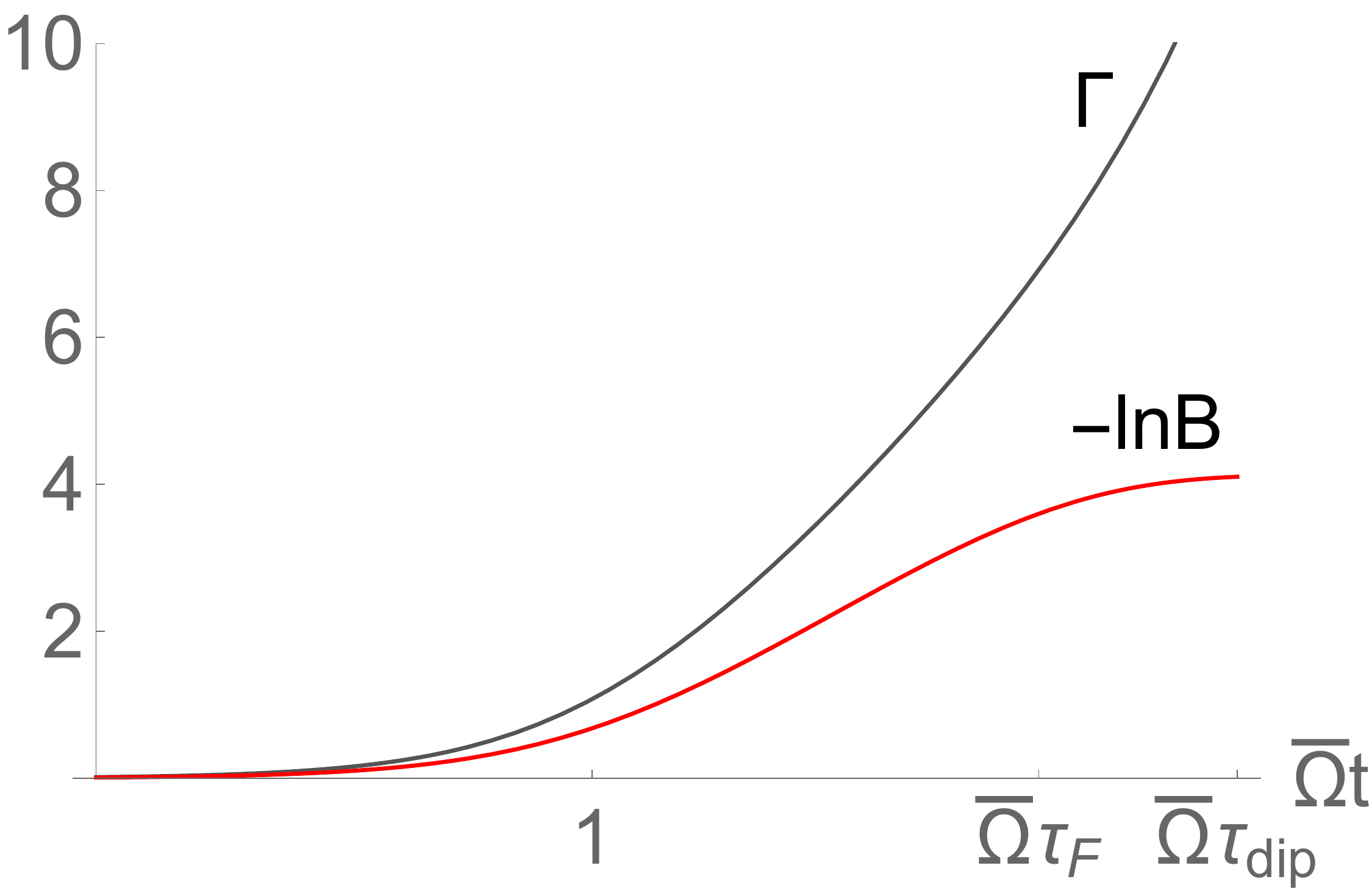}
\caption{(Color online). A sample plot of the decoherence damping factor $\Gamma_{\tvec p,\tvec p'}(t)$ (upper trace) and
the state fidelity  $-\log B^{mac}_{\tvec p,\tvec p'}(t)$ (lower trace) as a function of time, measured in the inverse
cutoff units $\bar\Omega^{-1}$ and plotted on a log scale.
The parameters for the plot are the following:  
$\bar\Omega = 10^{15}s^{-1}$, $\alpha = 10^5$, $\delta p_0/(m_0 c) = 5 \times 10^{-2}$, $\hbar \bar\Omega/(k_B T) = 4 \times 10^2$,
the unobserved portion $\Omega_{unob}$ is given by $0\leq\theta\leq \pi/4$, and the observed macrofraction size is $\Delta\Omega_0/(4\pi)=0.05$.
This gives $\tau_{dip} = 20 \bar\Omega^{-1}$ and $\tau_F \approx 7.95 \bar\Omega^{-1}$.
}\label{dec}
\end{figure}

Let us now introduce and study so called macrofractions of the field  \cite{Z2003,DDZ2004,Zureksfera,RZ2011, sfera}. 
We divide the monitored directions $\Omega_{obs}$ into patches $\Omega_{mac}$, each containing a large enough number of
modes to justify the continuum limit. The collection of all modes within $\Omega_{mac}$ with a fixed polarization defines 
a macrofraction with a given polarization.
Such a coarse-graining of the observed portion of the field may correspond e.g. 
to an array of (polarization-sensitive, wide-band) detectors; see Fig.~\ref{fig1}. In the box quantization, a state of a macrofraction
can be formally written as: 
\be
\varrho^{(j)}_{mac}(t;\tvec p)\equiv\bigotimes_{\tvec k/k\in\Omega_{mac}}\bigotimes_{\omega_{\tvec k}\lesssim\bar\Omega}\varrho_{\tvec k,j}(t;\tvec p).\label{rhomac}
\ee
We are interested in the mixed state fidelity
$B[\varrho^{(j)}_{mac}(t;\tvec p),\varrho^{(j)}_{mac}(t;\tvec p')]\equiv B^{mac}_{\tvec p,\tvec p',j}(t)$
for a fixed polarization $j$. 
Since $B(\varrho^{\otimes n},\sigma^{\otimes n})$ factorizes with respect to the tensor product, 
$B^{mac}_{\tvec p,\tvec p',j}(t)$ is a product taken over the macrofraction of the terms (\ref{Bmic}). 
Passing to the continuum limit and imposing the cut-off, we obtain:
\ben
\label{B}
&&\log B^{mac}_{\tvec p,\tvec p',j}(t) = -\frac{\alpha}{\pi(m_0c)^2} 
\int \frac{d\omega}{\omega} e^{-\frac{\omega}{\bar\Omega}}\tanh\left(\frac{\beta \omega}{2} \right)\nonumber\\
&&\times\int_{\Omega_{mac}}\frac{d\Omega_{\tvec k}}{4\pi}\left(\tsym\epsilon_{\tvec k,j}\cdot\Delta\tvec p\right)^2
\frac{1- \cos[(\omega-\tvec k\cdot\tvec v_0)t]}{[1-\tvec k\cdot\tvec v_0/(kc)]^2}
\label{logB}
\een
We are particularly interested in information content of small macrofractions, described by a small angle $\Delta\Omega_0$ 
centered around some $\tvec k_0$ (see Fig.~\ref{fig1}). It correspond to an almost point-like, from the macroscopic point of view, detector \cite{fluid}. 
We may then approximate $\int_{\Delta\Omega_0} d\Omega_{\tvec k}f(\tvec k)\approx f(\tvec k_0)\Delta\Omega_0$
and the remaining frequency integral can be calculated for $k_BT\ll\hbar\bar\Omega$, yielding:
\ben
\log B^{mac}_{\tvec p,\tvec p',j}(t) =
&&-\frac{\alpha\Delta\Omega_{0}\left(\tsym\epsilon_{\tvec k_0,j}\cdot\Delta\tvec p\right)^2}{4\pi^2 (m_0c)^2}\times\label{Bmac}\\
&&\frac{1}{\nu^2}\log \left[\sqrt{1+\nu^2 \bar{\Omega}^2 t^2}  
\frac{\tanh \left( \nu t / \tau_F  \right)}{ \nu t / \tau_F} \right],\nonumber
\een
where 
\be
\nu \equiv [1 - \tvec k_0\cdot\bold{v}_0/(k_0c) ]\label{nu}
\ee
is the Doppler factor along the direction of $\tvec k_0$.
Let us compare the behavior of (\ref{Bmac}) with that of the decoherence factor. Performing the same
approximations as in (\ref{Gappr}) yields:
\ben
-\bigg[&\frac{\alpha\Delta\Omega_{0}\left(\tsym\epsilon_{\tvec k_0,j}\cdot\Delta\tvec p\right)^2}{4\pi^2 (m_0c)^2}&\bigg]^{-1}
\log B^{mac}_{\tvec p,\tvec p'}(t)\approx\\
&&\approx\left\{\begin{array}{l r} \frac{\bar\Omega^2t^2}{2}, &  t\ll \bar\Omega^{-1}\\
\frac{1}{\nu^2}\log(\nu\bar\Omega t), & \bar\Omega^{-1}\ll t\ll\tau_F\\
\frac{1}{\nu^2}\log \left(\bar\Omega\tau_F\right), & 
t\gg\tau_F.\nonumber\label{Bappr}
\end{array} \right.
\een
We see that modulo the geometric factor (controlled by the solid angle of the directions),
the behavior of distinguishability, as measured by the above state fidelity, and decoherence is the same up to 
$t\sim\tau_F$, i.e. during the dressing the field acquires information about the momentum at the similar rate
as it decoheres the charge.
Past this time, the decoherence factor keeps decreasing (\ref{Gappr}), but the state distinguishability stabilizes
at: 
\be\label{Bmin}
B^{mac}_{\infty}\sim\left[\bar\Omega\tau_F\right]^{-\alpha\left(\frac{\Delta p}{m_0c}\right)^2}.
\ee
The reason is that the cut-off limits the energy, available for the displacement (\ref{Up}) of the initial thermal state
during the evolution. Since this displacement encodes the momentum data into the field, the cut-off puts a fundamental limit
on the accuracy with which the information about the momentum can be imprinted in and extracted from the thermal field \cite{bandwidth}
It is worth stressing that in our setup 
this is a thermal effect---for the field initially in the vacuum state, $B^{mac}$
decays without a limit, as follows from (\ref{Bmac}) with $T=0$. 
The accuracy is determined by
(\ref{Bmin}) and depends, among the others, on the
ratio of the cut-off and the thermal energies as $\bar\Omega\tau_F=\hbar\bar\Omega/(k_BT)$. 
The latter is small in the low energy regime considered here, however looking at the exponent in (\ref{Bmin})
a similar remark as after (\ref{Gappr}) applies:
the momentum difference to be discriminated cannot be arbitrarily small and a very strong coupling, $\alpha\gg 1$ is required for 
state fidelity to be small. This can be achieved, along with a vanishing decoherence factor,
as shown in Fig.~\ref{dec}.  

 For convenience, let us summarize the different time behaviors from Eqs. (\ref{Gappr}) and (\ref{Bappr}) in the following table:
\begin{widetext}
\begin{center}
    \begin{tabular}{ | c | c | c | }
    \hline
    Time-scale & $-$ Log of the modulus of decoherence factor  & $-$ Log of the state fidelity  \\ \hline
     \begin{tabular}{c c} $t\ll \bar\Omega^{-1}$ & \\ (time-dependent dressing)\end{tabular} & $F_0(\Delta \tvec{p})\frac{\bar\Omega^2t^2}{2}$ & $f(\Delta \tvec{p})\frac{\bar\Omega^2t^2}{2}$  \\ \hline
     \begin{tabular}{c c}$\bar\Omega^{-1}\ll t\ll\tau_F$ \\ (vacuum decoherence) \end{tabular} & $\log (\bar\Omega t) \left[F_0(\Delta \tvec{p})+\frac{v_0}{c}F_1(\Delta \tvec{p})\right]-\frac{v_0}{2c}F_1(\Delta \tvec{p})$ 
& $\frac{f(\Delta \tvec{p})}{\nu^2}\log(\nu\bar\Omega t)$  \\ \hline
     \begin{tabular}{c c}$t\gg\tau_F$ \\ (thermal decoherence) \end{tabular} & $\frac{t}{\tau_F}\left[F_0(\Delta \tvec{p})+\frac{v_0}{2c}F_1(\Delta \tvec{p})\right]+\left[F_0(\Delta \tvec{p})+\frac{v_0}{c}F_1(\Delta \tvec{p})\right]\log (\bar\Omega\tau_F)$ 
& $\frac{f(\Delta \tvec{p})}{\nu^2}\log \left(\bar\Omega\tau_F\right)$\\
    \hline
    \end{tabular}
\end{center}
Where $f(\Delta \tvec{p})\equiv \frac{\alpha\Delta\Omega_{0}\left(\tsym\epsilon_{\tvec k_0,j}\cdot\Delta\tvec p\right)^2}{4\pi^2 (m_0c)^2}$, $\nu$ is the Doppler factor given by (\ref{nu}), and $F_0(\Delta \tvec{p})$, $F_1(\Delta \tvec{p})$ by
(\ref{F0},\ref{F1}).
\end{widetext}

It implies that in the discussed parameter regime, the partially traced state (\ref{rhoobs}) approaches, 
so called, spectrum broadcast structure \cite{sfera,pra} and by the results of \cite{pra} provides a form of objectivization
of the charge momentum. Let us elaborate on that. SBS is defined as the following 
maximally correlated, classical-classical \cite{CC,qbreak} state (cf. \cite{ZQZ2009,RZZ2012,ZZ2013}):
\be\label{SBS}
\varrho=\sum_ip_i\ket i\bra i\otimes \varrho^{(1)}_i\otimes\cdots\otimes\varrho^{(M)}_i,
\ee  
where $\ket i$ is some basis (called pointer basis) in the system space, $p_i$ are probabilities and states $\varrho_i^{(m)}$, $m=1,\dots,M$,
have vanishing state fidelity for different $i$'s, $B[\varrho_i^{(m)},\varrho_{i'\ne i}^{(m)}]=0$ for all $m$. It has 
an important property that measuring the supports of $\varrho_i^{(m)}$ all the observers $m=1,\dots,M$ obtain the same index $i$
with the same probabilities $p_i$ in perfect correlation with the state of the system $\ket i$ and without disturbing (after forgetting the results)
the whole state $\varrho_{S:F_{obs}}$. In this sense the information about the state of the system is redundantly encoded in the
environment and can be extracted without perturbation. This, in turn, is at the core of what we perceive as objectivity \cite{ZurekNature,pra}.
Returning to the studied situation, vanishing of the decoherence factor and the state fidelities  imply \cite{sfera} that past $\tau_F$
the state (\ref{rhoobs})  is approximately of the form (somewhat abusing the notation and using 
the continuous distribution for $\tvec p$):  
\ben
\varrho_{S:F_{obs}}(t)&\approx& \int d^3 p \left|\langle\tvec p|\psi_{0S}\rangle\right|^2\ket{\tvec{p}}\bra{\tvec{p}}\otimes\\ \label{rhoobsass}
&&\bigotimes_j \left[\varrho^{(j)}_{mac_{\tvec k_0}}(t;\tvec p)\otimes\varrho^{(j)}_{mac_{\tvec k_1}}(t;\tvec p)\otimes\cdots\right],\nonumber
\een
where directions $\tvec k_0,\tvec k_1,\dots$ define the macrofractions into which the observed radiation is divided
and their states $\varrho^{(j)}_{mac_{\tvec k}}(t;\tvec p)$ have small state fidelities (\ref{Bmin}) for different momenta. 
Thus, although various types of quantum correlations, including entanglement, are produced during the evolution,
the ones that survive after the sufficiently long time, the partial loss of the field, and the coarse-graining are only of the SBS type.
However, although formally resembling an SBS, there is a key difference between (\ref{rhoobsass}) and (\ref{SBS}) and the structures 
encountered so far \cite{sfera,qbm,photonics}
(apart from the limit on the accuracy (\ref{Bmin})). 
By (\ref{Up}, \ref{rhomic},\ref{rhomac}) what is in fact encoded in each $\varrho^{(j)}_{mac_{\tvec k}}(t;\tvec p)$ is  the momentum component
$\tsym\epsilon_{\tvec k, j}\cdot\tvec p$ along the average macrofraction polarization vector corresponding to polarization $j$.
Thus, each macrofraction  carries in general different information about the same quantity $\tvec p$.
This situation resembles seeing different pieces of the same object. However, picking two different macrofractions, centered around
$\tvec k_0,\tvec k_1$ which are not antipodal, it is possible to choose three linearly independent polarizations $\tsym \epsilon_1,\tsym \epsilon_2,\tsym \epsilon_3$. 
Then, $\tvec p$ can be reconstructed from $\tsym \epsilon_i\cdot \tvec p$ using the Gramm matrix  
$G_{rs}\equiv \tsym \epsilon_r\cdot\tsym\epsilon_s$: $\tvec p=\sum G^{-1}_{rs}(\tsym\epsilon_s\cdot\tvec p)\tsym\epsilon_s$. 
In other words, any triple of polarization macrofractions in (\ref{rhoobsass}) with linearly independent
polarization vectors encodes almost perfect information about the charge momentum $\tvec p$.
If we now imagine that the observed "celestial sphere" $\Omega_{obs}$ can be divided
into a very large number of infinitesimal macrofractions $\Delta\Omega$, 
then it is clear that the information about $\tvec p$ is encoded with a huge redundancy in the field.
Moreover, it is available to multiple observers
without disturbing the state of the system (modulo the finite accuracy discussed above \cite{macro}). In this sense, the field in the studied regime 
provides an objectivization of the charge momentum. 

\section{Concluding remarks}
Our studies may be viewed as a step towards a more fundamental re-derivation, on the level of QED, 
of the results on objectivity \cite{Zureksfera,sfera} in the celebrated phenomenological model of decoherence due to environmental scattering \cite{JoosZeh}.
However, due to the used dipole approximation what becomes objective here is the momentum rather than the position.
In the context of a free charge, this approximation is the biggest limitation and a natural direction would be to go beyond it.
Another perspective would be systems with internal degrees of freedom, e.g. qubit models within QED \cite{Sowinski}.

 Finally, since we are explicitly including a part of the environment in the description, it may seem that we are dealing 
with a non Markovian evolution, where the role of the environment cannot be simplified to the usual Markovian generator. 
This is is not necessarily so - the reason for including the environment in the present description is to study information content of the environment
and not because its presence cannot be described in simple terms. The relation between Spectrum Broadcast Structures
and properly defined non-Markovianity has been studied in \cite{Lampo} (cf. \cite{Maniscalco}). There seems to be no obvious connection between the two,
at least in the context of the spin-boson model.

We would like to thank R. and P. Horodecki, D. Chru\'sci\'nski, K. Rza\.zewski, I. Bia\l{}ynicki-Birula and especially J. Karwowski for discussions. We acknowledge the financial support of 
the John Templeton Foundation through the grant ID \#56033.



\appendix 
\section{Derivation of the decoherence factor}
Here we present derivation leading to eq. (\ref{logD}) in the main text. It is a generalization of derivation presented in \cite{BCP2006} taking into account that in the present case only a portion $F_{unob}$ of the  field modes is unobserved.  From eqs. (\ref{eq:decf}), (\ref{Up}), we have 
\ben
&&- \Gamma_{\tvec{p}, \tvec{p}'}(t) =  \log |D_{\tvec{p}, \tvec{p}'}(t)| = \\ \nonumber && \log \left|  tr\left[\hat{U}_{F_{unob}}(t;\tvec{p}) \varrho_{0F_{unob}} \hat{U}_{F_{unob}}(t;\tvec{p}')^{\dagger}\right]\right| = \\ &&
\log \left|  tr \left[ \hat{D}\Big( \sum_{\tvec{k}_{unob},j} C_{\tvec{k}}\Delta \tvec{p} \cdot \tsym{\epsilon}_{\tvec{k},j}  \alpha_{\tvec{k}}(t)\Big) \varrho_{0F_{unob}} \right] \right|. \nonumber 
\een
If the environment is initially in a thermal state $\varrho_{0F}=\exp(-\beta \hat H_F)/Z(\beta)$ one finds \cite{BPbook,BCP2006,qbm}:
\ben
\label{app:decf}
&&- \Gamma_{\tvec{p}, \tvec{p}'}(t) = \\  \nonumber && \frac{1}{2}  \sum_{\tvec{k}_{unob},j} \left| C_{\tvec{k}} \alpha_{\tvec{k}}(t)\right|^2 \left|\Delta \tvec{p} \cdot \tsym{\epsilon}_{\tvec{k},j}   \right|^2 \coth \left( \frac{\beta \omega_k}{2}\right),
\een
with
\ben
\left| C_{\tvec{k}} \alpha_{\tvec{k}}(t)\right|^2= \frac{2 \pi q^2}{\hbar \omega_\tvec{k}^3 m_0^2 V} \frac{1-\cos(\omega_\tvec{k} t(1-\tvec{k}\cdot \tvec{v}_0/(kc)))}{(1 - \tvec{k} \cdot \tvec{v}_0/(kc))^2}. \nonumber
\een
Subsequently, we assume that the number of field modes in the unobserved fraction is large enough to pass to the continuum limit i.e. the sum over modes is replaced by an integral
\ben
&&\sum_{\tvec{k}}\to V\int_{F_{unob}} \frac{d^3 k}{(2 \pi)^3}=\frac{V}{2 \pi^2}\int_0^{\infty}  k^2 dk  \int_{\Omega_{unob}} \frac{d \Omega_\tvec{k}}{4 \pi} = \nonumber \\ && \frac{V}{2 \pi^2 c^3}\int_0^{\infty}   \omega^2 d \omega  \int_{\Omega_{unob}} \frac{d \Omega_\tvec{k}}{4 \pi}.
\een
In the above expression, the unobserved field modes $\tvec{k} \in F_{unob}$ are expressed in terms of spherical coordinates: The wave vector length $k= \omega/c$ and an angle $\Omega_{unob}$ of the unobserved directions (see Fig.~\ref{fig1} in the main text). Please note that the appearance of the speed of light $c$ here is a result of the dispersion relation $\omega=kc$. The Eq. (\ref{app:decf}) takes a form:
\ben
\label{eq: decfactd}
&&- \Gamma_{\tvec{p}, \tvec{p}'}(t) = \\  \nonumber
&&\frac{\alpha/ \pi}{(m_0 c)^2} \int_0^{\infty}\frac{d \omega}{\omega} \int_{\Omega_{unob}}  \frac{d \Omega_{\tvec{k}}}{4 \pi}   e^{-\frac{\omega}{\bar \Omega}} \coth  \left(\frac{\beta \omega}{2}\right) \nonumber \\ && \frac{1-\cos(\omega t(1-\tvec{k}\cdot \tvec{v}_0/(kc)))}{(1 - \tvec{k} \cdot \tvec{v}_0/(kc))^2} \sum_j \left|\Delta \tvec{p} \cdot \tsym{\epsilon}_{\tvec{k},j}   \right|^2,
\een
where additionally the cut-off was introduced.
The next step is to expand the fraction under integral in a series with respect to $\frac{\tvec{k}/k\cdot \tvec{v}_0 }{c} \equiv  \frac{v_0}{c}\cos \theta_{\tsym{k}}$
\ben
\label{eq:series}
&&\frac{1-\cos(\omega t(1-(v_0/c)\cos \theta_{\tsym{k}})}{(1 - (v_0/c)\cos \theta_{\tsym{k}})^2} \approx \\&& \nonumber (1-\cos(\omega t)) \left(1+2\frac{v_0}{c}\cos \theta_{\tsym{k}} + 3 \left(\frac{v_0}{c}\cos \theta_{\tsym{k}}\right)^2 
\right) \\&& \nonumber - \omega t \sin (\omega t)\frac{v_0}{c}\cos \theta_{\tsym{k}} \left( 1+ 2 \frac{v_0}{c}\cos \theta_{\tsym{k}} \right)  + \\
&& \nonumber (\omega t)^2 \cos (\omega t) \left(\frac{v_0}{c}\cos \theta_{\tsym{k}} \right)^2 + O\left(\frac{v_0^3}{c^3}\right).
\een 
Using the identity for polarization vectors
\ben
\sum_j  \tsym{\epsilon}^{n}_{\tsym{k},j}\tsym{\epsilon}^{m}_{\tsym{k},j} = \delta_{mn} - \tvec{k}_n \tvec{k}_m/k^2
\een
one easily establish that
\ben
\label{eq:pol}
\left|\Delta \tvec{p} \cdot \tsym{\epsilon}_{\tvec{k},j} \right|^2 = \sum_{ij}\Delta \tvec{p}_i\Delta \tvec{p}_j(\delta_{ij}-k_ik_j/k^2) \equiv \Delta \tvec{p}^2_{\perp\tvec k}
\een  
Inserting eqs (\ref{eq:series}, \ref{eq:pol}) into eq. (\ref{eq: decfactd}) leads to
\ben
\label{eq:appgamma}
&&\frac{\pi}{\alpha} \Gamma_{\tvec{p}, \tvec{p}'}(t) =\left[F_0(\Delta \tvec{p})+\frac{v_0}{c}F_1(\Delta \tvec{p})\right] \times \\&&
\int_0^{\infty} \frac{d \omega}{\omega} e^{-\frac{\omega}{\bar \Omega}} \coth  \left(\frac{\beta \omega}{2}\right)(1-\cos(\omega t))\nonumber \\ 
&&-\frac{v_0}{2c}F_1(\Delta \tvec{p})t \int_0^{\infty} d \omega  e^{-\frac{\omega}{\bar \Omega}} \coth  \left(\frac{\beta \omega}{2}\right)\sin(\omega t)\nonumber \\&& \nonumber + O\left(\frac{v_0^2}{c^2}\right),
\een
where
\ben
	&&F_0(\Delta \tvec{p})\equiv \frac{1}{(m_0 c)^{2}}\int_{\Omega_{unob}}\frac{d\Omega_{\tvec k}}{4\pi}\Delta \tvec{p}^2_{\perp\tvec k},\\
	&&F_1(\Delta \tvec{p})\equiv\frac{2}{(m_0 c)^{2}}\int_{\Omega_{unob}}\frac{d\Omega_{\tvec k}}{4\pi}\cos\theta_\tvec k\Delta \tvec{p}^2_{\perp\tvec k}.
	\een
The frequency integrals are split into vacuum and thermal contributions:
\ben
&&\int_0^{\infty} \frac{d \omega}{\omega}  e^{-\frac{\omega}{\bar \Omega}} \coth \left(\frac{\beta \omega}{2}\right)(1-\cos(\omega t)) = \Gamma^{vac}_1 + \Gamma^{th}_1 \nonumber \\ \\ \nonumber && \Gamma^{vac}_1 \equiv \int_0^{\infty} \frac{d \omega}{\omega}  e^{-\frac{\omega}{\bar \Omega}}(1-\cos(\omega t)) \\ \nonumber && \Gamma^{th}_1 \equiv   \int_0^{\infty} \frac{d \omega}{\omega}  e^{-\frac{\omega}{\bar \Omega}} \left[\coth  \left(\frac{\beta \omega}{2}\right)-1\right](1-\cos(\omega t)) \\
&& \int_0^{\infty} d \omega  e^{-\frac{\omega}{\bar \Omega}} \coth  \left(\frac{\beta \omega}{2}\right)\sin(\omega t) = \Gamma^{vac}_2 + \Gamma^{th}_2  \\&&  \nonumber \Gamma_2^{vac} \equiv \int_0^{\infty} d \omega  e^{-\frac{\omega}{\bar \Omega}} \sin(\omega t)  \\ \nonumber && \Gamma_2^{th} \equiv \int_0^{\infty} d \omega  e^{-\frac{\omega}{\bar \Omega}}\left[\coth  \left(\frac{\beta \omega}{2}\right)-1\right]\sin(\omega t)    
\een 
Evaluation of vacuum contributions is straightforward:
\ben
\label{eq:decv1}
\Gamma_1^{vac} = \frac{1}{2}\log \left[1+(\bar \Omega t)^2\right]\\
\label{eq:decv2}
\Gamma_2^{vac} = \frac{(\bar \Omega t)^2}{1+(\bar \Omega t)^2}.
\een
To arrive at close formulas for thermal contribution, in both cases we need to assume that the energy scale set by cut-off is much larger than the thermal energy i.e. $k_B T \ll \hbar \Omega$. Under this assumption one finds
\ben
&&\Gamma_1^{th} = \log \left[ \frac{\sinh\left(t/\tau_F\right)}{t/\tau_F}\right] \\
&&\Gamma_2^{th} = -\frac{1}{1+ \bar\Omega^2 t^2}.
\een
Combining the above expressions with eq. (\ref{eq:appgamma}) allows to arrive at eq. (\ref{logD}) of the main text. 	

Let us now briefly discuss the fidelity calculation. To arrive at Eq. (\ref{Bmic}), we used the derivation presented in \cite{qbm}. 
Subsequently we approximate the angular integral $\int_{\Delta\Omega_0} d\Omega_{\tvec k}f(\tvec k)\approx f(\tvec k_0)\Delta\Omega_0$ and split the frequency integral as
	\ben
\frac{1}{\nu^2}\int \frac{d\omega}{\omega} e^{-\frac{\omega}{\bar\Omega}}\tanh\left(\frac{\beta \omega}{2} \right)
\left(1- \cos(\nu t) \right) = \frac{1}{\nu^2}\left(B^{vac} + B^{th}\right),\nonumber \\
\een
where $\nu \equiv [1 - \tvec k_0\cdot\bold{v}_0/(k_0c) ]$.
The vacuum part is the same as for decoherence factor (eqs. (\ref{eq:decv1}, \ref{eq:decv2}))
	 \ben
	&&B^{vac} = \Gamma_1^{vac},
	 \een
	  whereas the thermal integral reads
\ben
	&&B^{th} =   \log \left[\frac{\tanh(\nu t/\tau_F)}{\nu t/ \tau_F}\right].
\een 

\section{Inclusion of higher order relativistic terms}\label{rel_cor}
Here we show that decoherence and fidelity are always two orders of magnitude higher than the Hamiltonian in a formal $1/c$ expansion.
This however is not a result of the relativistic effects per se, but rather of the continuum limit and the dispersion relation for light.
We will show it by taking into an account the first relativistic correction to the Hamiltonian (\ref{H}).
We start with the relativistic Hamiltonian. 
\ben
 \hat H= m_0c^2 \sqrt{1+ \frac{\hat{\boldsymbol{\pi}}^2}{m_0^2c^2}} -m_0c^2 + \sum_{\tvec{k},j} \hbar \omega_\tvec{k} \hat{a}^\dagger_{\tvec{k},j} \hat{a}_{\tvec{k},j},
\een
with canonical momentum 
\ben
\hat{\boldsymbol{\pi}} = \hat{\tvec{p}}-\frac{q}{c}\hat{\tvec{A}}(\hat{\tvec{r}}) 
\een
Expanding the square root up to $1/c^2$ we get
\ben
&&\hat H = \frac{\hat{\tvec{p}}^2}{2m_0} 
- \frac{\hat{\tvec{p}}^4}{8m_0^3c^2} +  \sum_{\tvec{k},j} \hbar \omega_\tvec{k} \hat{a}^\dagger_{\tvec{k},j} \hat{a}_{\tvec{k},j} -\frac{q}{m_0c}\hat{\tvec{p}} \cdot \hat{\tvec{A}}(\hat{\tvec{r}}) + \nonumber  \\ &&\frac{q}{4m_0^3c^3} \left( \hat{\tvec{p}}^2 \hat{\tvec{p}} \cdot \hat{\tvec{A}}(\hat{\tvec{r}}) +  \hat{\tvec{p}} \cdot \hat{\tvec{A}}(\hat{\tvec{r}}) \hat{\tvec{p}}^2 \right) + O\left(\hat{\tvec{A}}^2  (\hat{\tvec{r}})\right) 
\een
Subsequently we neglect terms proportional to $\hat{\tvec{A}}(\hat{\tvec{r}}(t))^2$ and use moving dipol approximation so that
\ben
&&\hat H \approx \frac{\hat{\tvec{p}}^2}{2m_0} 
- \frac{\hat{\tvec{p}}^4}{8m_0^3c^2} +  \sum_{\tvec{k},j} \hbar \omega_\tvec{k} \hat{a}^\dagger_{\tvec{k},j} \hat{a}_{\tvec{k},j} \label{Hc2}\\ && -\frac{q}{m_0c}\hat{\tvec{p}} \cdot \hat{\tvec{A}}(\tvec{r}(t)) +  \frac{q}{2m_0^3c^3}  \hat{\tvec{p}}^2 \hat{\tvec{p}} \cdot \hat{\tvec{A}}(\tvec{r}(t)) \nonumber   
\een
The interaction Hamiltonian in the interaction picture is
\ben
&&\hat H^I = \\ \nonumber && \int d \tvec{p} \left| \tvec{p} \left\rangle \right\langle \tvec{p} \right| \otimes  \left(-\frac{q}{m_0c}+\frac{q}{2m_0^3c^3  }\tvec{p}^2\right) \tvec{p} \cdot \hat{\tvec{A}}(\tvec{r}(t)).
\een
Therefore, the evolution operator can be written as
\ben\label{U}
\hat{U}^{I}_{S:F}(t) = \int d^3 p \ket{\tvec{p}}\bra{\tvec{p}} \otimes \hat{U}^{I}_{F}(t;\tvec{p}), 
\een
where
\ben
\hat{U}^{I}_{F}(t;\tvec{p})\equiv &&e^{i \sum_{\tvec{k},j} C_{\tvec{k}} \left(-1+\frac{1}{2m_0^2c^2}\tvec{p}^2\right)\tvec{p} \cdot \tsym{\epsilon}_{\tvec{k},j}\xi_\tvec{k}(t)}\times\\ \nonumber
&&\times\hat{D}\Big(\left(-1+\frac{1}{2m_0^2c^2}\tvec{p}^2\right) \sum_{\tvec{k},j} C_{\tvec{k}}\tvec{p} \cdot \tsym{\epsilon}_{\tvec{k},j}  \alpha_{\tvec{k}}(t)\Big).
\een
Repeating calculations of the previous section we find
\ben
\label{eq:relexqdec}
&&- \Gamma_{\tvec{p}, \tvec{p}'}^{(2)}(t) = \\  \nonumber && \frac{1}{2}  \sum_{\tvec{k}_{unob},j} \left| C_{\tvec{k}} \alpha_{\tvec{k}}(t)\right|^2 \left|\left[\Delta \tvec{p} + \frac{\tvec{p}^2 \tvec{p} - \tvec{p}'^2 \tvec{p}' }{2m_0^2c^2}\right] \cdot \tsym{\epsilon}_{\tvec{k},j}   \right|^2 \times \\ \nonumber &&\coth \left( \frac{\beta \omega_k}{2}\right),
\een
Proceeding as previously one arrives at
\ben
\label{eq:releqdecf}
&&- \Gamma_{\tvec{p}, \tvec{p}'}^{(2)}(t) = - \Gamma_{\tvec{p}, \tvec{p}'}^{(1)}(t) + \frac{\alpha/\pi}{(m_0c)^4} \left(I_1^{vac}+I_1^{th}\right) \nonumber \times \\ && \int_{\Omega_{unob}}\frac{d\Omega_{\tvec k}}{4\pi} \sum_j \left(\Delta \tvec{p} \cdot \tsym{\epsilon}_{\tvec{k},j}\right) \left[ \left(\tvec{p}^2 \tvec{p} - \tvec{p}'^2  \tvec{p}' \right)\cdot \tsym{\epsilon}_{\tvec{k},j}\right] \nonumber \\ && +
\frac{v_0^2}{c^2}\frac{\alpha/\pi}{(m_0c)^2}\left[2\left(I_1^{vac}+I_1^{th}\right)-2t\left(I_2^{vac}+I_2^{th}\right)\right. \nonumber \\&&  +\left. t^2 \left(I_3^{vac}+I_3^{th}\right)\right]  \times \int_{\Omega_{unob}}\frac{d\Omega_{\tvec k}}{4\pi} \cos^2 \theta_{\tvec{k}} \Delta \tvec{p}^2_{\perp\tvec k} ,
\een
where
\ben
&&\int_0^{\infty} d \omega  e^{-\frac{\omega}{\bar \Omega}} \coth  \left(\frac{\beta \omega}{2}\right) \omega \cos(\omega t) = I^{vac}_3 + I^{th}_3 \nonumber \\ \\ \nonumber && I^{vac}_3 \equiv \int_0^{\infty} d \omega  e^{-\frac{\omega}{\bar \Omega}}\omega \cos(\omega t) \\ \nonumber && I^{th}_3 \equiv   \int_0^{\infty} d \omega  e^{-\frac{\omega}{\bar \Omega}} \left[\coth  \left(\frac{\beta \omega}{2}\right)-1\right]\omega \cos(\omega t)
\een
The only change in calculation concerning fidelity will be that, starting from eq. (\ref{eq:relexqdec}) hyperbolic cotagnet will be replaced by hyperbolic tangent. This will result in different frequency integrals but will not change the conclusion regarding relativistic terms obtained in (\ref{eq:releqdecf}).

Hence we see that the logarithm of decoherence factor and fidelity  is formally of the fourth order in $1/c$
and is thus two orders higher  than the Hamiltonian (\ref{Hc2}). This is a general characteristic of this calculation: The effect will be formally two 
orders higher than the Hamiltonian. The root of this lies simply in the passage to the continuum limit and the use of the dispersion relation.


\begin{references}
\bibitem{nota2} State fidelity is the same in the interaction and Schr\"odinger pictures.
\bibitem{fluid} This is similar to the description of fluids: one
considers portions of the fluid which macroscopically
can be viewed as points but microscopically still contain a large enough amount
of particles to define e.g. the temperature.
\bibitem{bandwidth} In the context of information extraction, the
introduction of the cut-off in (\ref{logB}) may be given a natural
interpretation of the maximum
bandwidth of the detectors.
\end{references}
\end{document}